**An *Ab Initio* Full Potential Fully Relativistic Study of Atomic Carbon, Nitrogen, and Oxygen Chemisorption on the (111) Surface of δ-Pu**

Raymond Atta-Fynn and Asok K. Ray*

*Physics Department, University of Texas at Arlington, Arlington, Texas 76019*

*akr@uta.edu.

**Abstract**

First principles total energy calculations within the framework of generalized gradient approximation to density functional theory have been performed for atomic carbon, nitrogen, and oxygen chemisorptions on the (111) surface of δ-Pu. The full-potential all-electron linearized augmented plane wave plus local orbitals method with the Perdew-Burke-Ernzerhof (PBE) exchange-correlation functional has been employed . Chemisorption energies have been optimized with respect to the distance of the adatom from the Pu surface for four adsorption sites, namely the top, bridge, hollow fcc, and hollow hcp sites, the adlayer structure corresponding to a coverage of 0.50 of a monolayer in all cases. Computations were carried out at two theoretical levels, one without spin-orbit coupling (NSOC) and one with spin-orbit coupling (SOC). For NSOC calculations, the hollow fcc adsorption site was found to be the most stable site for C and N with chemisorption energies of 6.272 eV and 6.504 eV respectively, while the hollow hcp adsorption site was found to be the most stable site for O with chemisorption energy of 8.025 eV. For SOC calculations, the hollow fcc adsorption site was found to be the most stable site in all cases with chemisorption energies for C, N, and O being 6.539 eV, 6.714 eV, and 8.2 eV respectively. The respective distances of the C, N, and O adatoms from the surface were found to be 1.16 Å, 1.08 Å, and 1.25 Å. Our calculations indicate that SOC has negligible effect on the chemisorption geometries but energies with SOC are more stable than the cases with NSOC within a range of 0.05 to 0.27 eV. The work function and net magnetic moments respectively increased and decreased in all cases upon chemisorption compared with the bare δ-Pu (111) surface. The partial charges inside the muffin-tins, difference charge density distributions, and the local density of states have been used to analyze the Pu-adatom bond interactions.

## 1. Introduction

Considerable theoretical efforts have been devoted in recent years to studying the electronic and geometric structures and related properties of surfaces to high accuracy. One of the many motivations for this burgeoning effort has been a desire to understand the detailed mechanisms that lead to surface corrosion in the presence of environmental gases; a problem that is not only scientifically and technologically challenging but also environmentally important. Such efforts are particularly important for systems like the actinides for which experimental work is relatively difficult to perform due to material problems and toxicity. As is known, the actinides are characterized by a gradual filling of the 5$f$-electron shell with the degree of localization increasing with the atomic number Z along the last series of the periodic table and the increasing prominence of relativistic effects. [1-5] Narrower 5$f$ bands, with properties intermediate between those of localized 4$f$ and delocalized 3$d$ orbitals, near the Fermi level, compared to 4$d$ and 5$d$ bands in transition elements, is believed to be responsible for the exotic structure of actinides at ambient condition.[6-7]

The manmade plutonium metal (Pu) is located at the boundary between the light actinides (Th to Np) consisting of delocalized 5$f$ electrons and the heavy actinides (Am to Lw) consisting of localized 5$f$ electrons.[8-13] The face-centered cubic (fcc) δ-Pu is technologically important because it is highly ductile and this property makes it convenient for engineering applications.[14] This phase is usually stable in the temperature range 593-736 K; however, it can be stabilized at room temperature by small additions of alloying metals like Al and Ga.[15-17] In spite of detailed studies, δ-Pu is not well understood theoretically. Different theoretical approaches have yielded different degrees of success for δ-Pu.[18-40]

The primary focus of this work is to study the chemisorption and electronic structure of C, N, and O adatoms on δ-Pu (111) surface. The motivation for choosing this problem is a complete understanding of interactions and chemical reactivity of adatoms with Pu surfaces and the resulting surface electronic structure. As mentioned before, δ-Pu can be stabilized at room temperature by the addition of small amounts of impurities. Secondly, grazing-incidence photoemission studies, combined with the calculations of Eriksson *et al.*[41], suggest the existence of a small-moment δ-like surface on α-Pu. Using the linear combinations of Gaussian orbitals fitting function (LCGTO-FF) method as implemented in the suite of software GTOFF, Ray and Boettger[42] have also indicated the possibility of such a surface for a Pu monolayer. Recently, high-purity ultra-thin layers of Pu deposited on Mg were studied by X-ray photoelectron (XPS) and high-resolution valence band (UPS) spectroscopy by Gouder *et al.*[43], who found that the degree of delocalization of the 5*f* states depends in a very dramatic way on the layer thickness and that the itinerant character of the 5f states is gradually lost with reduced thickness, suggesting that the thinner films are δ-like. Finally, it may be possible to study 5*f* localization in Pu layers through adsorption on a series of carefully selected substrates in which case the adsorbed layers are more likely to be δ-like than α-like.

Experimental data[44] indicates that when Pu surface is exposed to molecular oxygen, oxygen is readily adsorbed by the metal surface. The oxygen molecule then dissociates into atomic oxygen, and combines with Pu to form a layer of oxide. Using the film linearized muffin tin orbitals (FLMTO) method, Eriksson *et al.*[41] have studied the electronic structure of hydrogen and oxygen chemisorbed on Pu. The Pu valence behavior was dominated by the 6*d* electrons, giving rise to significant hybridization with ligand valence electrons and significant covalency. There have also been studies in the literature of the bulk and surface electronic structures of PuO and water adsorption on the PuO$_2$ (110) surface by Wu and Ray[30], and a self-interaction corrected local spin density study by Petit *et al.*[45] of the electronic structure of PuO$_{2\pm x}$. Recently, detailed studies on adsorptions of H, H$_2$, O and O$_2$ on δ-Pu (111) and δ-Pu(100) have been carried out by Huda and Ray[40], using a scalar relativistic local basis density functional semi-core pseudo-potential (DSPP) approach using the non-spin-polarized and spin-polarized levels of theory within the generalized gradient approximation (GGA)[46] to density functional theory (DFT)[47] and the DMol$^3$ suite of programs.[48] We should mention here that, to the best of our knowledge, *no* theoretical work of C and N chemisorption on Pu surfaces has been reported in the literature.

## 2. Computational methodology

All calculations have been performed within the GGA to DFT with the Perdew-Burke-Ernzerhof (PBE) exchange-correlation functional[46]. The Kohn-Sham equations were solved using the all-electron full-potential linear augmented plane wave plus local basis (FP-LAPW+lo) method as implemented in the WIEN2k code[49]. This method makes no shape approximation to the potential or the electron density. Within the FP-LAPW+lo method, the unit cell is divided into non-overlapping muffin tins spheres and an interstitial region. Inside the muffin tin sphere of radius R$_{MT}$, the wave functions are expanded using radial functions (solution to the radial Schrödinger equation) times spherical harmonics up to $l_{max}^{wf}$, and the expansion of the potential inside the muffin tin spheres is carried out up to $l_{max}^{pot}$. The parameter $R_{MT}^{min} \times K_{MAX}$, where $R_{MT}^{min}$ is the smallest muffin tin spherical radius present in the system and $K_{MAX}$ is the truncation of the modulus of the reciprocal lattice vector, is used to determine the number of planes waves needed for the expansion of the wave function in the interstitial region while the parameter $G_{MAX}$ is used to truncate the plane wave expansion of the potential and density in the interstitial region. Here, we have used $R_{MT}(C) = R_{MT}(N) = R_{MT}(O) = 1.2$ Bohr, $R_{MT}(Pu) = 2.13$ Bohr, $l_{max}^{wf} = 10$, and $l_{max}^{pot} = 6$, $R_{MT}^{min} \times K_{MAX} = 7.35$ for $R_{MT}^{min} = 2.13$ Bohr, $R_{MT}^{min} \times K_{MAX} = 4.15$ for $R_{MT}^{min} = 1.2$ Bohr, and $G_{MAX} = 14$ Ry$^{1/2}$. Specifically, APW+lo basis is used to describe all $s, p, d,$ and $f$ ($l$=0, 1, 2, 3) states and LAPW basis for all higher angular momentum states up to $l_{max}^{wf} = 10$ in the expansion of the wave function in all cases. Furthermore, additional local orbitals (LO) were used to improve the description of semi-core states.

In the WIEN2k code, core states are treated at the fully relativistic level. Semi-core and valence states are treated at either the scalar relativistic level, i.e., no spin-orbit coupling (NSOC) or at the fully relativistic level, i.e., spin-orbit coupling (SOC) included. Spin-orbit interactions for semi-core and valence states are incorporated via a second variational procedure using the scalar relativistic eigenstates as basis[49], where all eigenstates with energies below the cutoff energy of 4.5 Ry were included, with the so-called $p_{1/2}$ extension[49], which accounts for the finite character of the wave function at the nucleus for the $p_{1/2}$ state. We considered both the NSOC and SOC levels of theory to investigate the effects of spin-orbit coupling on chemisorption energies.

The δ-Pu (111) surface is modeled by a supercell consisting of periodic 3-layer slabs with two atoms per surface unit cell, where periodic slabs are separated in the z-direction by vacuum regions of 60 Bohr thick. Our use of the 3-layer slab is justified by recent calculations δ-Pu surfaces which showed that surface properties converge within the first three layers.[50] Also, our recent DFT-GGA calculations on bulk and (111) surfaces of δ-Pu have shown that the lowest energy configurations correspond to anti-ferromagnetic (AFM) arrangement of electron spins.[40] We have therefore used an AFM configuration for our surface which consist of alternating ferromagnetic layers of up- or down-spin atoms with the axis of magnetization being [001]. All surfaces were constructed using the optimized AFM bulk theoretical lattice constant of 8.55 Bohr[51] which is a 2.4 % contraction of the experimental lattice constant. We hasten to point out that better agreement with the experimental lattice constant has been obtained by us in our previous calculations (1.1% contraction)[40] but here we had to choose a smaller $R_{MT}$ for Pu to avoid spheres from overlapping during chemisorption. For the sake of consistency and for comparative different atomic adsorption studies, we have used the bulk theoretical lattice constant for all surface calculations. Thus, *no* further surface relaxations and/or reconstructions of the surface have been taken into account primarily because of *computational costs* and therefore, in some sense, the results can be considered as *preliminary*. We do believe though that the qualitative *and* quantitative results reported here will *not* change significantly upon the inclusions of relaxations and/or reconstructions. Integrations in the Brillouin zone (BZ) have been performed using the special k-points sampling method with the temperature broadening of the Fermi surface by the Fermi distribution where a broadening parameter of 0.005 Ry has been used. This scheme avoids the instability originating from level crossings in the vicinity of the Fermi surface in metallic systems and also reduces the number of k-points necessary to calculate the total energy of metallic systems.[49,52] For the present work, 18 k-points in the irreducible part of the BZ were found to be sufficient. Self-consistency is achieved when the total energy variation from iteration to iteration converged to a 0.01 mRy accuracy or better.

To study adsorption on the Pu surface, a single adatom, corresponding to a surface coverage of 0.5 ML, was allowed to approach the surface from one side along four different symmetrical positions as shown in figure 1: i) top site(adatom is directly on top of a Pu atom); ii) bridge site (adatom is placed in the middle of two nearest neighbor Pu atoms); iii) hcp hollow site (adatom sees a Pu atom located on the layer directly below the surface); and iv) fcc hollow site (adatom sees a Pu atom two layers below the surface). The chemisorption energy $E_C$ is optimized with respect to the height R of the adatom above the surface. The chemisorption energy $E_C$ is given by:

$$E_C(R) = E(M) + E(X) - E(M+X),$$

where E(M) is the total energy of the bare metal slab, E(X) is the total energy of the isolated adatom, and E(M+X) is the total energy of the adatom adsorbed on the metal. To calculate the total energy of the adatom, the isolated atom was simulated in a large box of side 30 Bohr and at the $\Gamma$ k-point.

**3. Results and discussions**

Table 1 lists the adsorption energies and associated geometrical information of the C, N, and O atoms adsorbed on the (111) surface of δ-Pu. The differences between the NSOC and SOC chemisorption energies at each adsorption site, given by $\Delta E_C = E_C(SOC) - E_C(NSOC)$, are also listed. For C adsorption, the trends in the chemisorption energies at the NSOC level of theory are the same as those in the SOC case. The most stable site is the hollow fcc site (6.272 eV for the NSOC case, 6.539 eV for SOC case) closely followed by the hollow hcp site (6.150 eV for the NSOC case, 6.414 eV for the SOC case). The least favorable site is the top site (3.899 eV for the NSOC case, 3.985 eV in the SOC case) with the bridge adsorption site having an intermediate chemisorption energy (5.787 eV for NSOC case, 5.983 eV for the SOC case). The vertical height R of the C atom above the top layer clearly show that at the least stable top site, the adatom is furthest away from the surface (1.99 Å for the NSOC case, 1.95 Å for the SOC case) followed by the intermediately stable bridge site (1.34 Å for the NSOC case, 1.35 Å for the SOC case). The vertical height of the adatom from the surface layer is smallest at both the most stable hollow fcc site and the next stable hollow hcp site with a degeneracy at the NSOC level of theory (1.18 Å) and is almost degenerate at the SOC level of theory (1.16 Å at the hollow fcc site, 1.17 Å at the hollow hcp site). Hence, increasing stability at both the NSOC and SOC levels of theory implies decreasing vertical distance of the C adatom from the surface layer. Also increasing adatom coordination number implies increasing stability at both theoretical levels; top site in one-fold coordinated, bridge site is two-fold coordinated, hollow sites are three-fold coordinated. The Pu-C bond lengths listed in table 1 also indicate a relationship with the adatom coordination numbers, with the one-fold coordinated top site having the shortest bond and the three-fold hollow-sites having the longest bonds. Also, as expected, all chemisorption energies in the SOC case are more stable than the NSOC case. $\Delta E_C$ is maximum at the most stable hollow fcc site (0.267 eV) closely followed by the next stable hollow hcp adsorption site (0.264 eV), with the intermediately stable bridge adsorption site having an SOC-NSOC $\Delta E_C$ = 0.196 eV. The least stable top site has the smallest energy difference of 0.086 eV. Overall, inclusion of spin-orbit coupling increases the chemisorption energy but the effect on the adsorption geometry is negligible.

N adsorption on the δ-Pu (111) surface closely follows the results obtained for C but with slightly favorable chemisorption energies. The hollow fcc site (6.504 eV in the NSOC case, 6.714 eV in the SOC case) is more slightly stable than the hollow hcp site (6.400 eV in the NSOC case, 6.603 eV in the SOC case). The bridge site is slightly less stable (6.105 eV in the NSOC case, 6.231 eV in the SOC case), with the top site again being the least stable (4.490 eV in the NSOC case, 4.544 eV in the SOC case) site. The trends in the height of N above the surface, bond lengths, coordination number in relation to stability is similar to the discussion for C above and can be inferred from table 1. The SOC-NSOC energy differences also follow exactly the trends observed for C.

The results for O adsorption is similar to C and N as far as trends in the chemisorption are concerned with the only exception being the order of increasing stability in the NSOC case. In this case, unlike the cases for C and N, the hollow hcp site is the most stable site (8.036 eV) closely followed by the hollow fcc site (8.025 eV) with the bridge site being intermediately stable (7.629 eV) and the top site again being the least stable (6.750 eV). For the SOC case, the order of decreasing stability is as follows: a degeneracy between the hollow fcc and hollow hcp sites (8.2 eV), followed by the bridge site (7.777 eV), and the top site (6.75 eV). The relationship between increasing stability and decreasing distance from the surface at both the NSOC and SOC theoretical levels is similar to that of C and N. The relationship between increasing stability and increasing coordination numbers and increasing Pu-O bond distance, which was observed for C and N, can also be observed here. Also, the SOC chemisorption energies are lower than the NSOC chemisorption energies. The only notable difference is the SOC and NSOC chemisorption energy differences at a given site, which, unlike the case for C and N, is fairly constant. The NSOC chemisorption energies and vertical height of the O adatom from the surface can be compared to a previous scalar relativistic *ab initio* localized basis semi-core pseudo-potential DFT calculations[40] for the same supercell. The O chemisorption energies reported in Ref. [40] were 6.14 eV, 7.238 eV, and 7.217 eV respectively for the top, bridge, and hollow fcc sites. Compared to our corresponding NSOC results of 6.75 eV, 7.629 eV, and 8.025 eV, for the top, bridge and hollow fcc sites respectively, we observe a chemisorption energy difference of 0.61 eV, 0.391 eV, and 0.808 eV between the current and previous chemisorption energies. The differences are quite significant and possibly stem from the different computational methodologies. However the distances of the O atom from the surface, which are 1.84 Å for the top site, 1.46 Å for the bridge site, and 1.33 Å for the fcc hollow site agree fairly well with our respective values of 1.84 Å for the top site, 1.34 Å for the bridge site, and 1.23 Å for the fcc hollow site, which confirms the fact that geometric predictions of structures do not necessarily require a highly accurate level of theory in contrast to energies. In general, we find the relations $r_d(\text{hollow}) < r_d(\text{bridge}) < r_d(\text{top})$ and $E_C(\text{hollow}) > E_C(\text{bridge}) > E_C(\text{top})$ to be true.

We would like to comment on difference in adsorption energies between sites in relation to energy barriers to diffusion across the surface. Diffusion of an adatom from the most stable hollow site to the neighboring hollow site will proceed via the lowest energy pathway. In all cases, this will be via the bridge site. From table 1, the energy difference between from the most stable hollow site and bridge site for C, N, and O are respectively 0.49 eV (0.57 eV), 0.40 eV (0.48 eV), and 0.41 eV (0.42 eV) for the NSOC (SOC) cases. These energy differences suggest that the energy barriers to diffusion across the surface are pretty much uniform for all the adatoms.

In table 2, the adsorbate-induced work function changes with respect to the clean metal surface, given by $\Delta\Phi = \Phi^{adatom/Pu} - \Phi^{Pu}$, are listed for the NSOC and SOC levels of theory for each adsorbate and each adsorption site. We observe for each adatom and each theoretical level that high chemisorption energies usually correspond to low work function shifts. In fact, the changes in the work functions are largest at the least preferred top sites and lowest at the most preferred hollow sites. The trends in the work function agrees well with previous works.[40,41] The work function shifts can be understood in terms of the surface dipoles arising due to the partial transfer of electrons from the Pu surface to the adsorbates since the electronegativities of all the adatoms are larger than that of Pu. The surface dipole moment μ (in Debye) and the work function shift ΔΦ (in eV) are linearly related by the Helmholtz equation $\Delta\Phi = 12\Pi\Theta\mu/A$, where A is the area in Å$^2$ per (1×1) surface unit cell and Θ is the adsorbate coverage in monolayers.[53] From the Helmholtz equation, we see that for each adsorbed adatom, μ is largest at the top site and lowest at the hollow site.

In table 3, the magnitude and alignment of the site projected spin magnetic moments for each Pu atom on each atomic layer, as well as the net spin magnetic moment per Pu atom at the SOC theoretical level are reported for the clean metal surface and the chemisorbed systems. NSOC moments follow a similar trend and are not reported here. Here $\mu_1, \mu_2$, and $\mu_3$ are respectively the spin moments for each of the two Pu atoms in each layer for the surface, middle, and bottom layers; $\mu_{int}$ and $\mu_{tot}$ are respectively the interstitial spin moment and net moment per Pu atom respectively. First, it is clearly evident from table 2 that the values of $\mu_2$ and $\mu_3$ in the chemisorbed cases is virtually the same as that of the clean metal surface, with the major changes in the spin magnetic moments occurring primarily in the top surface layer. This also justifies our use of a 3-layer slab. As a result, all discussions regarding the spin magnetic moments will be confined to the surface layer. For the top sites, we note reductions in the spin moment of the atom on top of which the adatom directly sits while the moment on the second atom remain practically unaltered compared to the clean metal. This leads to a reduction in the net spin magnetic moment per Pu atom. For the bridge sites, we see an equal reduction in the spin moments of each surface layer Pu atom since the adatom sits exactly between with the two Pu atoms, leading to a reduction in the net spin magnetic moment. For the hollow hcp sites, we see a slight reduction in the spin moment for the first Pu atom and a significant reduction in the spin moment for the second Pu atom. The case for the hollow fcc sites is exactly the same as the hcp hollows but with moments on the Pu atoms interchanged. Again, chemisorption leads to a reduction of the net spin moment for all hollow sites.

Due to the nature of the APW+lo basis, the electronic charges inside the muffin-tin spheres can be decomposed into contributions from different angular momentum channels. We refer to these charges as partial charges. In table 4 the partial charge contributions for C chemisorbed on the metal surface are reported. $Q_B$ is the partial charge inside a muffin tin sphere before adsorption, $Q_A$ partial charge inside a muffin tin sphere after adsorption and $\Delta Q = Q_A - Q_B$ is the difference in partial charges at a given adsorption site. A positive value of $\Delta Q$ indicates charge gain while a negative value indicates otherwise. The partial charges are reported at only the SOC level of theory for the valence *p* states of the adatom and the valence *d* and *f* of the Pu atoms. The NSOC partial charges exhibit a similar qualitative behavior and minor quantitative differences and are not reported here. Looking at the partial charges at each site in table 4, we see that compared to the partials charges of the clean slab little or no changes occur on the middle and bottom layers, while major changes occur on the surface layer. This is also true for the cases of N and O reported in tables 5 and 6 respectively. Hence, just like the spin magnetic moments, the discussions on the changes in the partial charges induced by chemisorption will be focused on only the surface layer. For the top site, the change in the partial charges occur for the *p* state of the C and the *d* and *f* states of the Pu atom on which the adatom sits. Looking at $\Delta Q$, we observe a gain of 0.12 e for the C *p* state, a gain of about 0.18 e for the Pu *d* state and a loss of about 0.16 e in Pu *f* state, suggesting significant C(2*p*)-Pu(5*f*)-Pu(6*d*) interactions. For the bridge site, we observe that the C *p* partial charges increase by about 0.21 e (which about twice that of the top site), while the Pu *d* partial charges for both atoms increase by 0.08 e and the Pu *f* partial charges decrease by 0.11 e. For the hcp site, we observe an increase of about 0.25 e for the C *p* partial charges. For the Pu *d* partial charges, one atom gains 0.05 e and the other atom gains 0.11 e, while for the Pu *f* partial charges, one atom loses 0.06 e and the other atom, 0.16 e. For the fcc site, the partial charge changes for the C *p* state is similar to the to that of the hcp site while for two surface Pu atoms the changes are the same but with the charges on the atoms interchanged.

The partial charges for N and O chemisorptions are reported in tables 5 and 6 respectively. From the values of ΔQ in tables 5 and 6, we observe the same qualitative trend as observed in table 4 for C chemisorption. However, it is worth noting that for each of the adatoms at a given site, the gain in partial $p$ partial charges in O is the greatest, followed, by N and C, in accordance with the order of decreasing electronegativities of the adatoms. Also, a general decrease in $f$ partial charges of the Pu atoms clearly signifies the participation of the Pu $5f$ electrons in bonding.

The partial charges discussed above were analyzed solely inside the muffin-tin, and this does not give us a complete picture of the nature of the bonds between the adatoms and the surface Pu atoms since it does not address the charge distribution in the interstitial region. To see any bonds that may have formed between the adatoms and the Pu atoms on the surface, we computed the difference charge density distribution for the O chemisorbed system, which gives us information about the nature of the chemical bonds formed as result of charge redistribution. We define the difference charge density Δn(r) as follows:

Δn(r) = n(O+Pu) – n(Pu) – n(O),

where n(O+Pu) is the total electron charge density of the O adatom adsorbed on the surface, n(Pu) is the total charge density of the bare Pu metal slab, and n(O) is the total charge density of the O adatom. In computing n(O) and n(Pu), the Pu and O atoms are kept fixed at exactly the same positions as they were in the chemisorbed systems. All densities reported here were computed in the plane passing through the adatom and the two surface Pu atoms. In figure 2, the difference charge density distribution is shown for each site. For the top site, we see clearly see charge accumulation around the O adatom and significant charge loss around the Pu atom on which the O atoms sits, implying that the bond has a strong ionic character. However, as one moves along the Pu-O bond, we see some charge accumulation along the part of the bond in the interstitial region, indicating that the bond has a small degree of covalent bonding. We also see some charge accumulation on the Pu atoms. This picture is clearly consistent with the changes in partial charges induced by chemisorption reported in table 6, where the O $p$ states gains charge, the Pu $f$ states loses charge and the Pu $d$ states gains charge. For the bridge site, the significant charge accumulation around the O adatom is evident with clear charge depletion around both Pu atoms as one moves from O to Pu along the Pu-O bonds. This suggests that bonding is primarily ionic. The difference electron charge density distributions for the hollow hcp and hollow fcc sites are qualitatively similar to that of the bridge site, clearly indicating strong ionic bonds. The difference charge density distributions for C and N show the same behavior and are not reported here.

We have also examined the local density of electron states (LDOS). This is obtained by decomposing the total density of the single particle Kohn-Sham eigenstates into contributions from each angular momentum channel $l$ of the constituent atoms inside the muffin tin sphere. We have reported the LDOS for only the SOC computation. In figure 3, the $f$ and $d$ LDOS curves for each layer of the clean Pu metal slab are shown. Because of the symmetry of the surface, the LDOS for each atom on a given layer is the same as the other so only LDOS plot of one atom is shown. Clearly we see peaks in the $f$ electron LDOS in the vicinity of the Fermi level, which might indicate *some* 5*f* electron localization. This assertion is in general agreement with published works in the literature.

In figure 4, we show the LDOS plots for the C adatom and the two surface Pu atoms. We note that the 5*f* DOS of the atom Pu1 on which the adatom sits is greatly modified, with a significant reduction of the DOS at the Fermi level (in comparison to the surface layer LDOS for the clean Pu slab in figure 3), which implies that the 5*f* electron participates in chemical bonding and are therefore delocalized. On the other hand, the 5*f* DOS of Pu2 is slightly modified, implying that its 5*f* electrons still primarily retain their localization. We also observe complete Pu(6*d*)-Pu1(5*f*)-C(2*p*) hybridizations. For the bridge site, we note a similar LDOS distribution for both Pu atoms. In comparison to the surface layer LDOS of the clean slab, we again observe a reduction in the 5*f* DOS at the Fermi level, indicating the participation of the 5*f* orbitals in bonding. Pu(6*d*)-Pu(5*f*)-C(2*p*) hybridization is also evident. The behavior of the LDOS plots for the hollow fcc and hollow hcp sites are qualitatively similar to that of bridge site. The LDOS of the atoms in the middle and bottom layers indicate no significant changes from those of the bare slab.

In figure 5, the LDOS plots for N chemisorbed systems are shown. For the top site, we see a significant reduction in the 5*f* DOS of Pu1 in the vicinity of the region below the Fermi level and slight modifications to the 5*f* DOS of Pu2. This indicates the strong 5*f* electron contribution from Pu1 to chemical bonding. Significant Pu(6*d*)-Pu1(5*f*)-N(2*p*) hybridizations can also be observed. For bridge, hollow hcp and hollow fcc sites, we see a significant reduction in the first peak of the 5*f* DOS (deep inside the valence band) in comparison to the LDOS for the surface layer of the clean Pu slab and a small reduction at the Fermi level, implying possibly the delocalization of some 5*f* electrons. However, unlike the top site, the hybridization of the electron states is dominated by Pu(6$d$)-N(2$p$) with a small admixture with the Pu 5$f$ states.

In figure 6, we depict the LDOS plots for the O chemisorption. The behavior of the LDOS at each site is qualitatively similar to that of C and N. For the top site, the hybridizations of the eigenstates is evident, with a significant reduction in the 5$f$ DOS of Pu1. We would like to mention that the Pu 5$f$ and O 2$p$ hybridization we have observed here has also been observed by Wu and Ray.[30] For the remaining sites, we observe significant hybridizations between the 6$d$ states of Pu and the 2$p$ states of O with a small admixture with Pu 5$f$ states. The reduction in the heights of the first peak of the 5$f$ DOS, possibly leading to the delocalization of some the 5$f$ electrons is observed as in the cases for C and N chemisorption.

## 4. Conclusions

In summary, we have used the generalized gradient approximation to density functional theory with the full potential LAPW+lo method to study chemisorption of C, N, and O atoms on the (111) surface of δ-Pu at two theoretical levels; one with no spin-orbit coupling (NSOC) and the other with spin-orbit coupling (SOC). Except for O which has the hollow hcp site as the most stable site at the NSOC theoretical level, the hollow fcc site was found to be the most preferred site in all cases with the adatom, in general, being closest to the surface. The inclusion of spin-orbit coupling lowers the chemisorption by 0.05-0.27 eV but the geometry of the chemisorbed system basically remains unaltered at both the NSOC and SOC theoretical levels. Work functions increased in all cases compared to the clean Pu surface, with the largest shift corresponding to the least preferred top site and lowest shifts corresponding to the hollow sites. Upon adsorption, the net magnetic moment of the chemisorbed system decreases in each case compared to the bare surface. Analysis of the partial charges of each atom confined within the muffin-tin spheres indicated that chemisorption takes place primarily on the uppermost layer. Partial charge inside the muffin-tin spheres and difference charge density plots clearly show that bonds between the surface Pu atom and the adatoms at each site is largely ionic in character. A study of the local density of states indicate that at the least favorable top site a significant reduction in the 5$f$ DOS at the Fermi level of the Pu atom on-top of which the adatoms sit implying partial delocalization of the 5$f$ electrons upon chemisorption. Significant Pu (6$d$)-Pu1(5$f$)-Adatom(2$p$) is observed for the least stable top site. For all other sites, Pu-adatom hybridizations is dominated by Pu 6$d$ and adatom 2$p$ states, with a significant reduction in the first peak of the 5$f$ DOS, indicating the delocalization of some of the 5$f$ electrons.

**Acknowledgements**

This work is supported by the Chemical Sciences, Geosciences and Biosciences Division, Office of Basic Energy Sciences, Office of Science, U. S. Department of Energy (Grant No. DE-FG02-03ER15409) and the Welch Foundation, Houston, Texas (Grant No. Y-1525).

Table 1: Chemisorption energies $E_c$, distances of the adatoms from the surface layer $R$, distances of the adatoms from the nearest neighbor Pu atom $D_{Pu-adatom}$ at both the NSOC and SOC levels of theory. $\Delta E_C = E_C(SOC) - E_C(NSOC)$ is the difference between the chemisorption energies at each adsorption site.

| Adatom | Site | NSOC | | | SOC | | | |
| --- | --- | --- | --- | --- | --- | --- | --- | --- |
| | | R (Å) | $E_C$ (eV) | $D_{Pu-Adatom}$ (Å) | R (Å) | $E_C$ (eV) | $D_{Pu-Adatom}$ (Å) | $\Delta E_C$ (eV) |
| Carbon | Top | 1.99 | 3.899 | 1.99 | 1.95 | 3.985 | 1.95 | 0.086 |
| | Bridge | 1.34 | 5.787 | 2.09 | 1.35 | 5.983 | 2.09 | 0.196 |
| | Hcp | 1.18 | 6.150 | 2.19 | 1.17 | 6.414 | 2.19 | 0.264 |
| | Fcc | 1.18 | 6.272 | 2.19 | 1.16 | 6.539 | 2.18 | 0.267 |
| Nitrogen | Top | 1.81 | 4.490 | 1.81 | 1.80 | 4.544 | 1.80 | 0.054 |
| | Bridge | 1.24 | 6.105 | 2.02 | 1.23 | 6.231 | 2.02 | 0.126 |
| | Hcp | 1.12 | 6.400 | 2.16 | 1.14 | 6.603 | 2.17 | 0.203 |
| | Fcc | 1.11 | 6.504 | 2.15 | 1.08 | 6.714 | 2.14 | 0.210 |
| Oxygen | Top | 1.84 | 6.750 | 1.84 | 1.84 | 6.906 | 1.84 | 0.156 |
| | Bridge | 1.34 | 7.629 | 2.09 | 1.31 | 7.777 | 2.07 | 0.148 |
| | Hcp | 1.27 | 8.036 | 2.24 | 1.22 | 8.200 | 2.21 | 0.163 |
| | Fcc | 1.23 | 8.025 | 2.22 | 1.25 | 8.200 | 2.23 | 0.175 |

Table 2: Changes in work functions $\Delta\Phi = \Phi^{adatom/Pu} - \Phi^{Pu}$ (in eV) at the NSOC and SOC levels of theory. $\Phi^{Pu}$ = 3.260 eV and 3.488 eV respectively at the NSOC and SOC levels of theory.

| Theory | Site | Carbon | Nitrogen | Oxygen |
|---|---|---|---|---|
| NSOC | Top | 1.648 | 1.839 | 1.531 |
|  | Bridge | 1.365 | 1.015 | 0.862 |
|  | Hcp | 1.155 | 0.920 | 0.835 |
|  | Fcc | 1.160 | 0.960 | 0.680 |
| SOC |  |  |  |  |
|  | Top | 1.439 | 1.529 | 1.358 |
|  | Bridge | 1.176 | 0.838 | 0.641 |
|  | Hcp | 0.964 | 0.803 | 0.541 |
|  | Fcc | 0.929 | 0.667 | 0.555 |

Table 3: $\mu_1$, $\mu_2$, $\mu_3$ are respectively the site projected spin magnetic moment for each Pu atom for the surface layer, middle layer and bottom layer. $\mu_{int}$ is the total spin magnetic moment in the interstitial region and $\mu_{tot}$ is the net (site + interstitial) magnetic moment per atom. Spin moments are quoted for SOC calculations.

|  | Site | $\mu_1$ ($\mu_B$) | $\mu_2$ ($\mu_B$) | $\mu_3$ ($\mu_B$) | $\mu_{int}$ ($\mu_B$) | $\mu_{tot}$ ($\mu_B$/Pu atom) |
|---|---|---|---|---|---|---|
| Bare Slab |  | 4.17, 4.17 | -3.82, -3.82 | 4.17, 4.17 | 2.97 | 2.00 |
| Carbon | Top | 3.46, 4.11 | -3.82, -3.81 | 4.15, 4.15 | 2.21 | 1.74 |
|  | Bridge | 3.62, 3.62 | -3.87, -3.84 | 4.16, 4.15 | 1.93 | 1.63 |
|  | Hcp | 3.86, 3.40 | -3.87, -3.84 | 4.16, 4.15 | 2.02 | 1.65 |
|  | Fcc | 3.38, 3.82 | -3.88, -3.84 | 4.16, 4.16 | 1.94 | 1.62 |
| Nitrogen | Top | 2.90, 4.15 | -3.80, -3.79 | 4.15, 4.15 | 2.02 | 1.63 |
|  | Bridge | 3.70, 3.70 | -3.87, -3.84 | 4.17, 4.16 | 2.13 | 1.69 |
|  | Hcp | 3.93, 3.60 | -3.86, -3.84 | 4.16, 4.16 | 2.23 | 1.73 |
|  | Fcc | 3.60, 3.92 | -3.86, -3.83 | 4.16, 4.15 | 2.21 | 1.73 |
| Oxygen | Top | 3.33, 4.19 | -3.82, -3.82 | 4.16, 4.16 | 2.27 | 1.75 |
|  | Bridge | 3.91, 3.91 | -3.85, -3.83 | 4.17, 4.16 | 2.35 | 1.80 |
|  | Hcp | 4.02, 3.84 | -3.85, -3.83 | 4.16, 4.16 | 2.39 | 1.82 |
|  | Fcc | 3.85, 4.04 | -3.85, -3.86 | 4.16, 4.15 | 2.39 | 1.81 |

Table 4: Partial charges inside muffin tin spheres before adsorption ($Q_B$), after adsorption ($Q_A$), and difference in partial charges $\Delta Q = Q_A - Q_B$ at the various adsorption sites for C at the SOC level of theory.

| Site | Atom/Layer | Partial charges in muffin-tin | | | | | | $\Delta Q = Q_A - Q_B$ | | |
| --- | --- | --- | --- | --- | --- | --- | --- | --- | --- | --- |
| | | Before adsorption $Q_B$ | | | After adsorption $Q_A$ | | | | | |
| | | C $p$ | Pu $d$ | Pu $f$ | C $p$ | Pu $d$ | Pu $f$ | C $p$ | Pu $d$ | Pu $f$ |
| Top | Carbon | 0.5926 | | | 0.7150 | | | 0.1224 | | |
| | Pu surface layer | | 0.4354 | 4.4298 | | 0.6118 | 4.2742 | | 0.1764 | -0.1556 |
| | | | 0.4354 | 4.4298 | | 0.4040 | 4.4457 | | -0.0314 | 0.0159 |
| | Pu middle layer | | 0.5023 | 4.3198 | | 0.4996 | 4.3238 | | -0.0027 | 0.0040 |
| | | | 0.5023 | 4.3198 | | 0.5023 | 4.3225 | | 0.0000 | 0.0027 |
| | Pu bottom layer | | 0.4354 | 4.4319 | | 0.4355 | 4.4300 | | 0.0001 | -0.0019 |
| | | | 0.4354 | 4.4319 | | 0.4355 | 4.4286 | | 0.0001 | -0.0033 |
| Bridge | Carbon | 0.5926 | | | 0.8066 | | | 0.2140 | | |
| | Pu surface layer | | 0.4354 | 4.4298 | | 0.5153 | 4.3182 | | 0.0799 | -0.1116 |
| | | | 0.4354 | 4.4298 | | 0.5166 | 4.3187 | | 0.0812 | -0.1111 |
| | Pu middle layer | | 0.5023 | 4.3198 | | 0.4980 | 4.3366 | | -0.0043 | 0.0168 |
| | | | 0.5023 | 4.3198 | | 0.4978 | 4.3351 | | -0.0045 | 0.0153 |
| | Pu bottom layer | | 0.4354 | 4.4319 | | 0.4368 | 4.4385 | | 0.0014 | 0.0066 |
| | | | 0.4354 | 4.4319 | | 0.4355 | 4.4371 | | 0.0001 | 0.0052 |
| hcp | Carbon | 0.5926 | | | 0.8383 | | | 0.2457 | | |
| | Pu surface layer | | 0.4354 | 4.4298 | | 0.4821 | 4.3683 | | 0.0467 | -0.0615 |
| | | | 0.4354 | 4.4298 | | 0.5408 | 4.2690 | | 0.1054 | -0.1608 |
| | Pu middle layer | | 0.5023 | 4.3198 | | 0.4995 | 4.3359 | | -0.0028 | 0.0161 |
| | | | 0.5023 | 4.3198 | | 0.4975 | 4.3389 | | -0.0048 | 0.0191 |
| | Pu bottom layer | | 0.4354 | 4.4319 | | 0.4375 | 4.4312 | | 0.0021 | -0.0007 |
| | | | 0.4354 | 4.4319 | | 0.4353 | 4.4282 | | -0.0001 | -0.0037 |
| Fcc | Carbon | 0.5926 | | | 0.8414 | | | 0.2488 | | |
| | Pu surface layer | | 0.4354 | 4.4298 | | 0.5493 | 4.2579 | | 0.1139 | -0.1719 |
| | | | 0.4354 | 4.4298 | | 0.4843 | 4.3591 | | 0.0489 | -0.0707 |
| | Pu upper layer | | 0.5023 | 4.3198 | | 0.4951 | 4.3411 | | -0.0072 | 0.0213 |
| | | | 0.5023 | 4.3198 | | 0.4974 | 4.3336 | | -0.0049 | 0.0138 |
| | Pu upper layer | | 0.4354 | 4.4319 | | 0.4369 | 4.4308 | | 0.0015 | -0.0011 |
| | | | 0.4354 | 4.4319 | | 0.4359 | 4.4347 | | 0.0005 | 0.0028 |

Table 5: Partial charges inside muffin tin spheres before adsorption ($Q_B$), after adsorption ($Q_A$), and difference in partial charges $\Delta Q = Q_A - Q_B$ at the various adsorption sites for N at the SOC level of theory.

| Site | Atom/Layer | Partial charges in muffin-tin | | | | | | $\Delta Q = Q_A - Q_B$ | | |
| --- | --- | --- | --- | --- | --- | --- | --- | --- | --- | --- |
| | | Before adsorption $Q_B$ | | | After adsorption $Q_A$ | | | | | |
| | | N $p$ | Pu $d$ | Pu $f$ | N $p$ | Pu $d$ | Pu $f$ | N $p$ | Pu $d$ | Pu $f$ |
| Top | Nitrogen | 1.2758 | | | 1.4736 | | | 0.1978 | | |
| | Pu surface layer | | 0.4354 | 4.4298 | | 0.6745 | 4.2538 | | 0.2391 | -0.1760 |
| | | | 0.4354 | 4.4298 | | 0.4026 | 4.4636 | | -0.0328 | 0.0338 |
| | Pu middle layer | | 0.5023 | 4.3198 | | 0.5005 | 4.3205 | | -0.0018 | 0.0007 |
| | | | 0.5023 | 4.3198 | | 0.5030 | 4.3185 | | 0.0007 | -0.0013 |
| | Pu bottom layer | | 0.4354 | 4.4319 | | 0.4355 | 4.4296 | | 0.0001 | -0.0023 |
| | | | 0.4354 | 4.4319 | | 0.4343 | 4.4360 | | -0.0011 | 0.0041 |
| Bridge | Nitrogen | 1.2758 | | | 1.5316 | | | 0.2558 | | |
| | Pu surface layer | | 0.4354 | 4.4298 | | 0.5046 | 4.3396 | | 0.0692 | -0.0902 |
| | | | 0.4354 | 4.4298 | | 0.5074 | 4.3362 | | 0.0720 | -0.0936 |
| | Pu middle layer | | 0.5023 | 4.3198 | | 0.5014 | 4.3328 | | -0.0009 | 0.0130 |
| | | | 0.5023 | 4.3198 | | 0.5012 | 4.3289 | | -0.0011 | 0.0091 |
| | Pu bottom layer | | 0.4354 | 4.4319 | | 0.4367 | 4.4358 | | 0.0013 | 0.0039 |
| | | | 0.4354 | 4.4319 | | 0.4354 | 4.4382 | | 0.0000 | 0.0063 |
| hcp | Nitrogen | 1.2758 | | | 1.5229 | | | 0.2471 | | |
| | Pu surface layer | | 0.4354 | 4.4298 | | 0.4646 | 4.3825 | | 0.0292 | -0.0473 |
| | | | 0.4354 | 4.4298 | | 0.5014 | 4.2970 | | 0.0660 | -0.1328 |
| | Pu middle layer | | 0.5023 | 4.3198 | | 0.5033 | 4.3319 | | 0.0010 | 0.0121 |
| | | | 0.5023 | 4.3198 | | 0.5020 | 4.3335 | | -0.0003 | 0.0137 |
| | Pu bottom layer | | 0.4354 | 4.4319 | | 0.4364 | 4.4297 | | 0.0010 | -0.0022 |
| | | | 0.4354 | 4.4319 | | 0.4353 | 4.4343 | | -0.0001 | 0.0024 |
| Fcc | Nitrogen | 1.2758 | | | 1.5402 | | | 0.2644 | | |
| | Pu surface layer | | 0.4354 | 4.4298 | | 0.5129 | 4.2911 | | 0.0775 | -0.1387 |
| | | | 0.4354 | 4.4298 | | 0.4711 | 4.3728 | | 0.0357 | -0.0570 |
| | Pu upper layer | | 0.5023 | 4.3198 | | 0.5000 | 4.3362 | | -0.0023 | 0.0164 |
| | | | 0.5023 | 4.3198 | | 0.5000 | 4.3285 | | -0.0023 | 0.0087 |
| | Pu upper layer | | 0.4354 | 4.4319 | | 0.4382 | 4.4329 | | 0.0028 | 0.0010 |
| | | | 0.4354 | 4.4319 | | 0.4362 | 4.4349 | | 0.0008 | 0.0030 |

Table 6: Partial charges inside muffin tin spheres before adsorption ($Q_B$), after adsorption ($Q_A$), and difference in partial charges $\Delta Q = Q_A - Q_B$ at the various adsorption sites for O at the SOC level of theory.

| Site | Atom/Layer | Partial charges in muffin-tin | | | | | | $\Delta Q = Q_A - Q_B$ | | |
| --- | --- | --- | --- | --- | --- | --- | --- | --- | --- | --- |
| | | Before adsorption $Q_B$ | | | After adsorption $Q_A$ | | | | | |
| | | O $p$ | Pu $d$ | Pu $f$ | O $p$ | Pu $d$ | Pu $f$ | O $p$ | Pu $d$ | Pu $f$ |
| Top | Oxygen | 2.1075 | | | 2.3336 | | | 0.2261 | | |
| | Pu surface layer | | 0.4354 | 4.4298 | | 0.6043 | 4.2434 | | 0.1689 | -0.1864 |
| | | | 0.4354 | 4.4298 | | 0.4085 | 4.4693 | | -0.0269 | 0.0395 |
| | Pu middle layer | | 0.5023 | 4.3198 | | 0.4980 | 4.3290 | | -0.0043 | 0.0092 |
| | | | 0.5023 | 4.3198 | | 0.5018 | 4.3248 | | -0.0005 | 0.0050 |
| | Pu bottom layer | | 0.4354 | 4.4319 | | 0.4360 | 4.4325 | | 0.0006 | 0.0006 |
| | | | 0.4354 | 4.4319 | | 0.4341 | 4.4352 | | -0.0013 | 0.0033 |
| Bridge | Oxygen | 2.1075 | | | 2.3727 | | | 0.2652 | | |
| | Pu surface layer | | 0.4354 | 4.4298 | | 0.4533 | 4.3864 | | 0.0179 | -0.0434 |
| | | | 0.4354 | 4.4298 | | 0.4551 | 4.3834 | | 0.0197 | -0.0464 |
| | Pu middle layer | | 0.5023 | 4.3198 | | 0.5023 | 4.3277 | | 0.0000 | 0.0079 |
| | | | 0.5023 | 4.3198 | | 0.5020 | 4.3228 | | -0.0003 | 0.0030 |
| | Pu bottom layer | | 0.4354 | 4.4319 | | 0.4399 | 4.4338 | | 0.0045 | 0.0019 |
| | | | 0.4354 | 4.4319 | | 0.4387 | 4.4367 | | 0.0033 | 0.0048 |
| hcp | Oxygen | 2.1075 | | | 2.3735 | | | 0.2660 | | |
| | Pu surface layer | | 0.4354 | 4.4298 | | 0.4427 | 4.4161 | | 0.0073 | -0.0137 |
| | | | 0.4354 | 4.4298 | | 0.4461 | 4.3733 | | 0.0107 | -0.0565 |
| | Pu middle layer | | 0.5023 | 4.3198 | | 0.5040 | 4.3189 | | 0.0017 | -0.0009 |
| | | | 0.5023 | 4.3198 | | 0.5039 | 4.3179 | | 0.0016 | -0.0019 |
| | Pu bottom layer | | 0.4354 | 4.4319 | | 0.4366 | 4.4368 | | 0.0012 | 0.0049 |
| | | | 0.4354 | 4.4319 | | 0.4350 | 4.4398 | | -0.0004 | 0.0079 |
| Fcc | Oxygen | 2.1075 | | | 2.3662 | | | 0.2587 | | |
| | Pu surface layer | | 0.4354 | 4.4298 | | 0.4368 | 4.3633 | | 0.0014 | -0.0665 |
| | | | 0.4354 | 4.4298 | | 0.4370 | 4.4036 | | 0.0016 | -0.0262 |
| | Pu upper layer | | 0.5023 | 4.3198 | | 0.5036 | 4.3342 | | 0.0013 | 0.0144 |
| | | | 0.5023 | 4.3198 | | 0.5013 | 4.3269 | | -0.0010 | 0.0071 |
| | Pu upper layer | | 0.4354 | 4.4319 | | 0.4375 | 4.4320 | | 0.0021 | 0.0001 |
| | | | 0.4354 | 4.4319 | | 0.4355 | 4.4353 | | 0.0001 | 0.0034 |

**Top view**

(a) Top 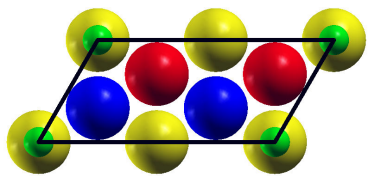
(b) Bridge 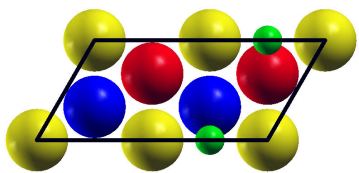
(c) Hcp 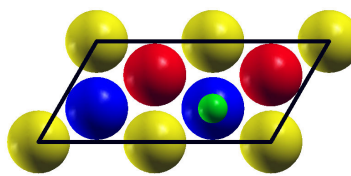
(d) Fcc 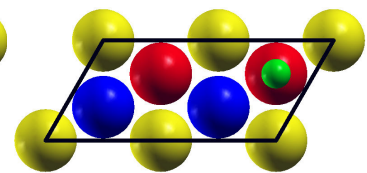

**Side view**

(a) Top 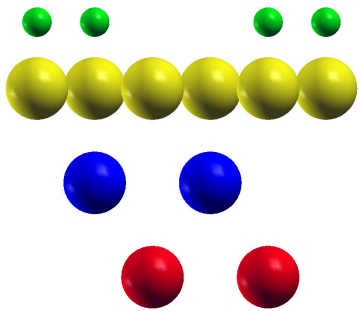
(b) Bridge 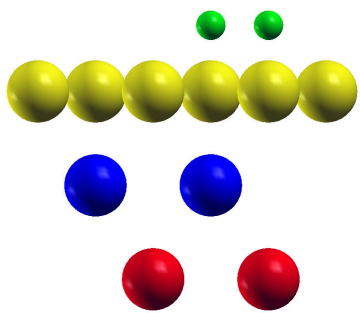
(c) Hcp 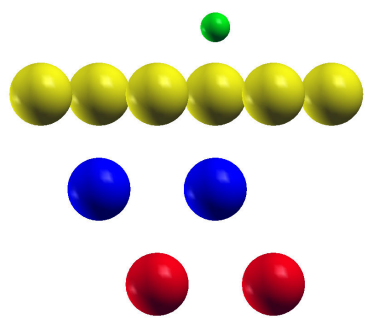
(d) Fcc 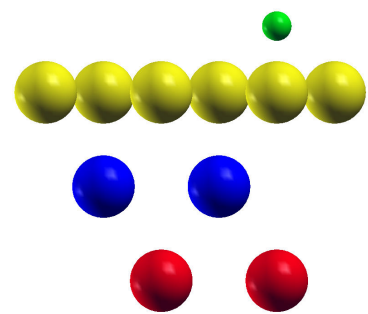

Figure 1 (color online): Top and side view illustrations of the four high-symmetry adsorption sites for the 3-layer δ-Pu(111) slab with a 0.5 ML adlayer coverage. Atoms are colored to distinguish between the layers. Surface, middle and top layers are colored gold, blue and red respectively. Adatom is colored green.

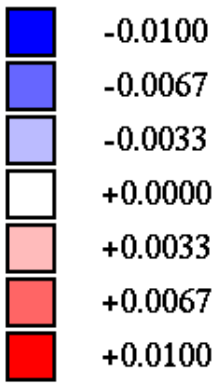

**(a) Top**  **(a) Bridge**

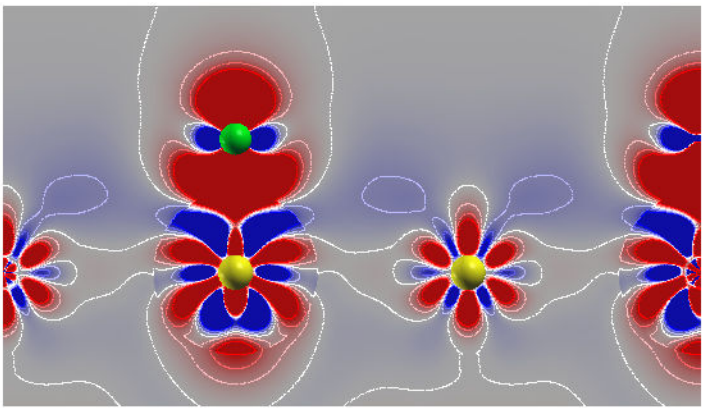  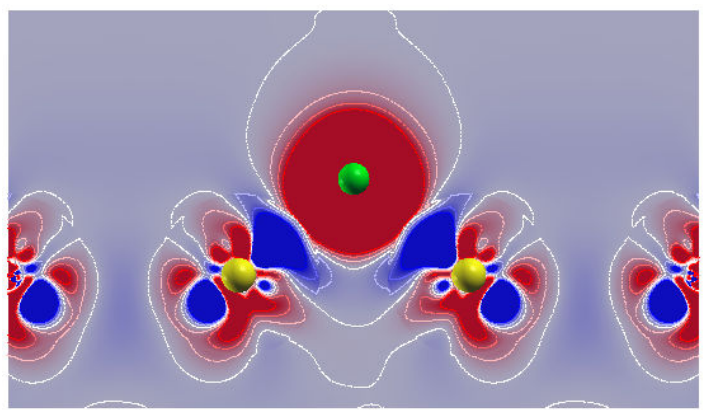

**(c) Fcc**  **(c) Hcp**

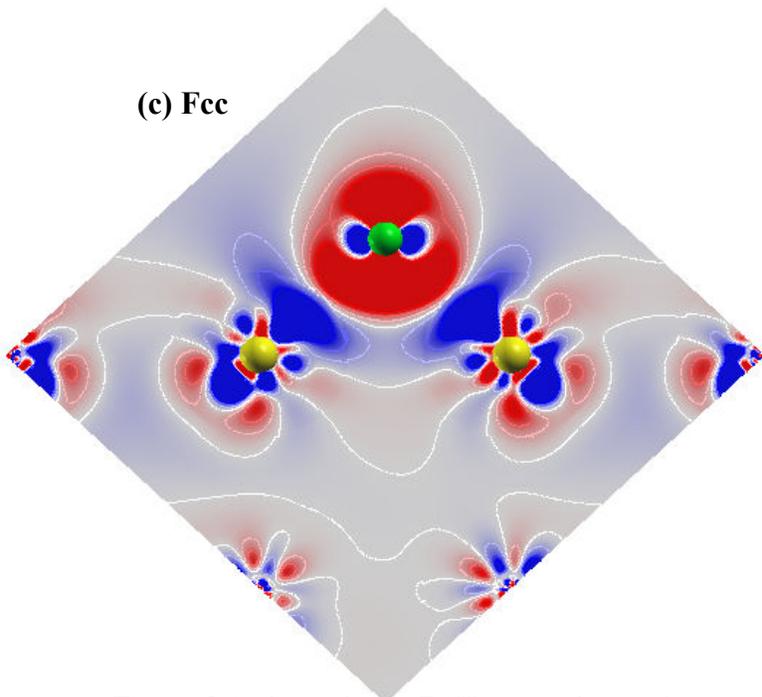  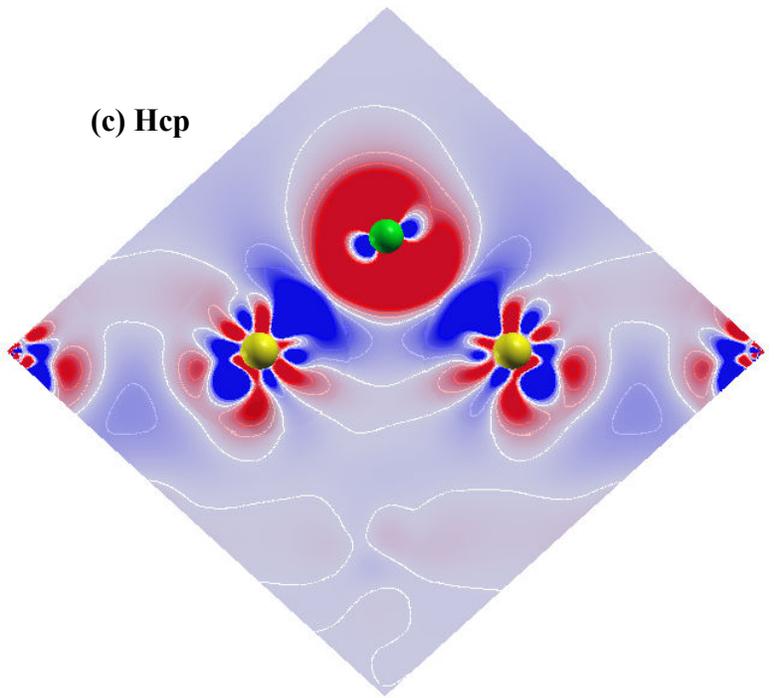

Figure 2 (color online): Difference charge density distributions Δn(r) for O on δ-Pu(111) surface at 0.50 monolayer coverage. O atom is colored green and Pu atoms are colored gold. The scale used for coloring is shown at the top. Red (positive) denotes regions of charge accumulation and blue (negative) denotes regions of charge depletion.

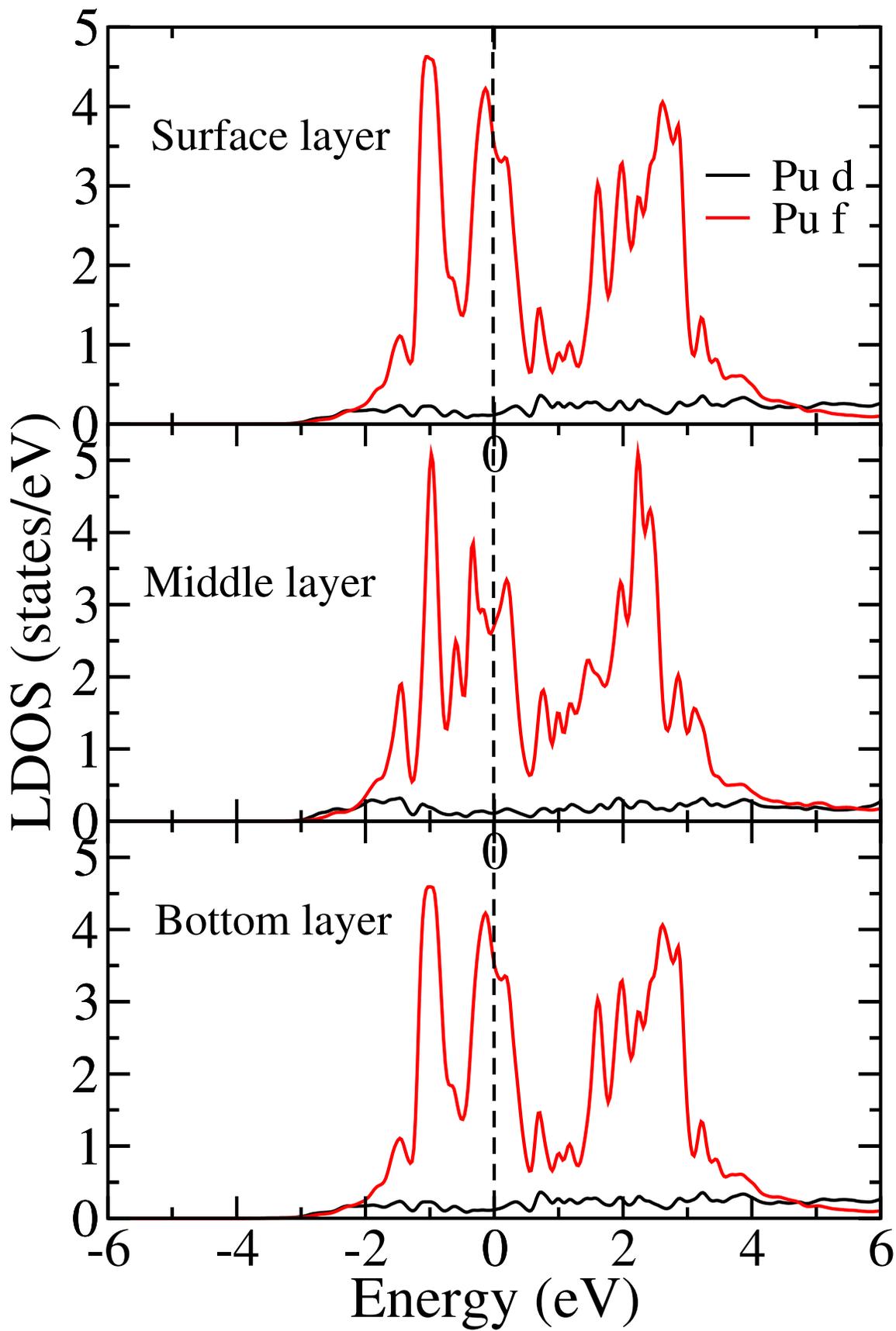

Figure 3 (color online): *d* and *f* LDOS curves inside the muffin-tins for each layer of the bare δ-Pu(111) slab. Vertical line through E=0 is the Fermi level. LDOS correspond to calculations with SOC.

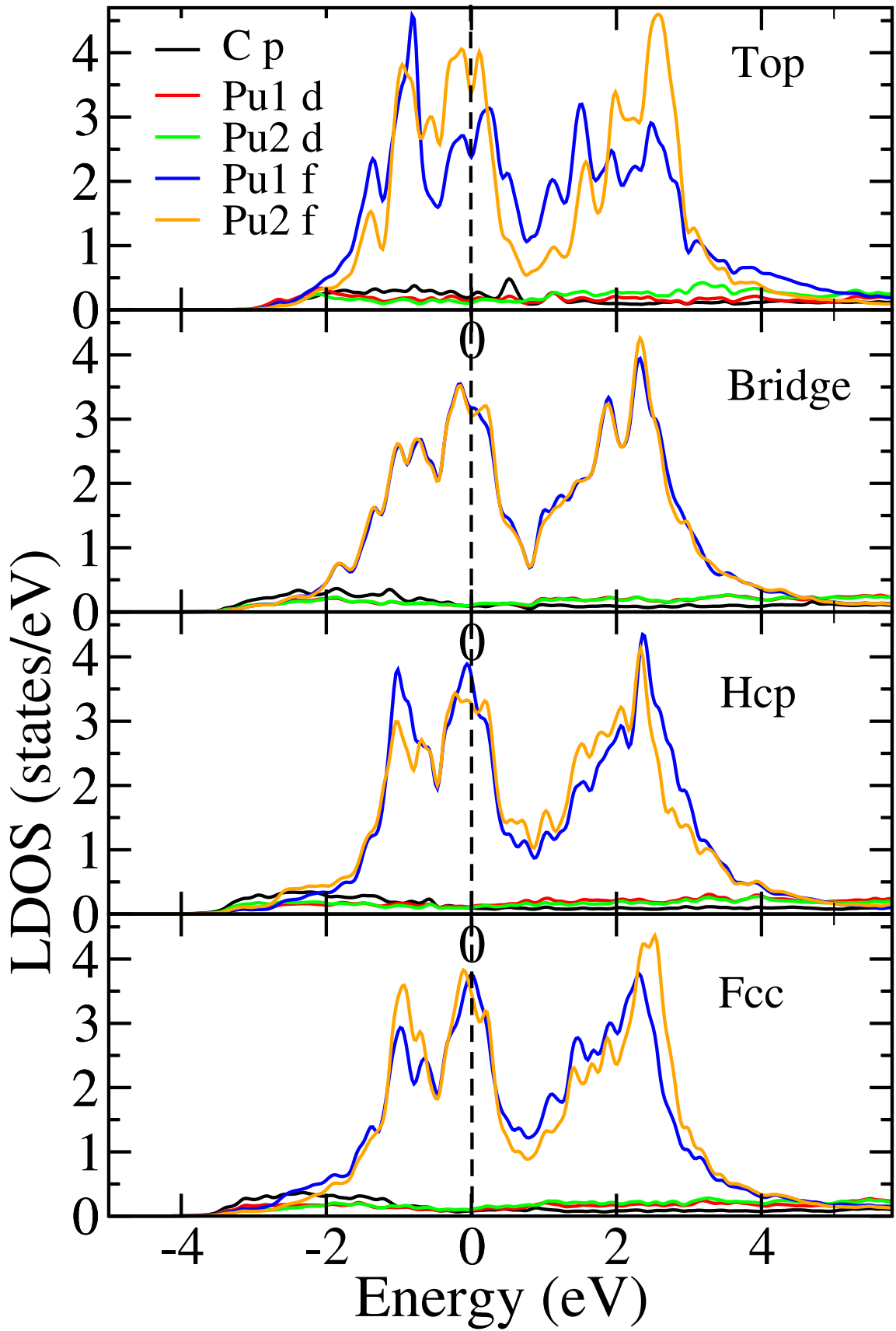

Figure 4 (color online): *d* and *f* LDOS curves inside the muffin-tins for the Pu atoms on the surface layer and *p* LDOS curves for C adatom. At the top site, C sits on-top of Pu1. Vertical line through E=0 is the Fermi level. LDOS correspond to calculations with SOC.

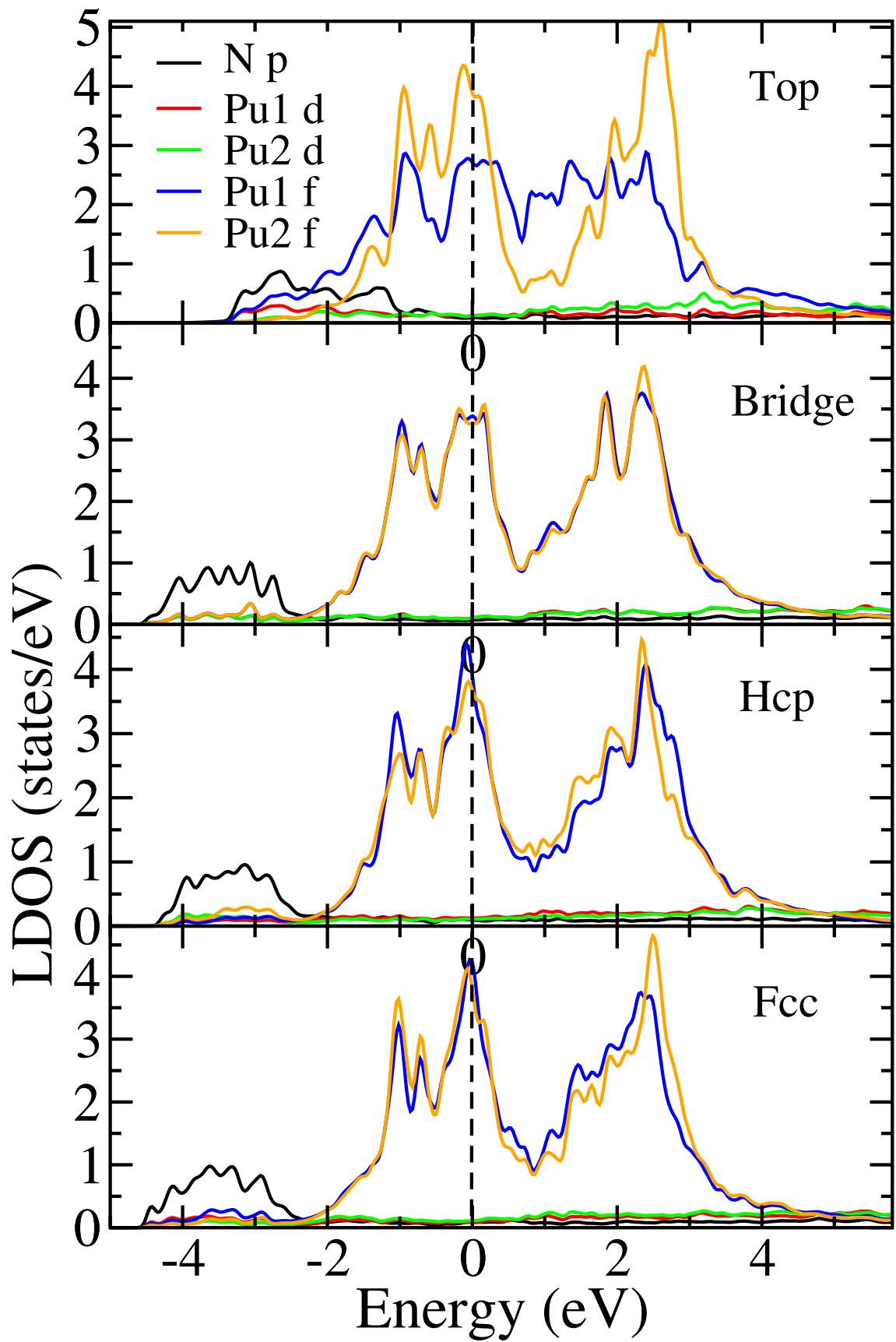

Figure 5 (color online): *d* and *f* LDOS curves inside the muffin-tins for the Pu atoms on the surface layer and *p* LDOS curves for N adatom. At the top site, N sits on-top of Pu1. Vertical line through E=0 is the Fermi level. LDOS correspond to calculations with SOC.

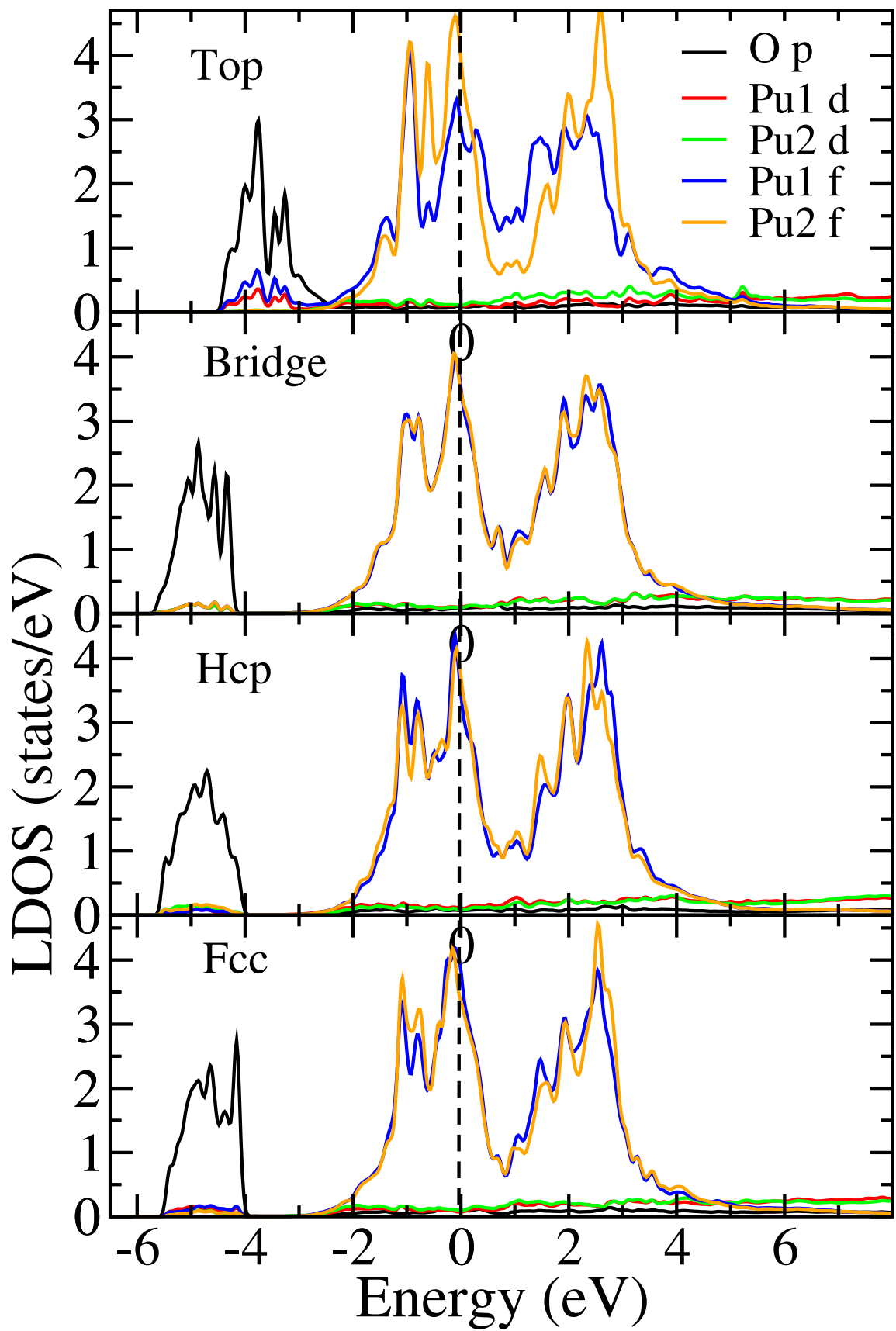

Figure 6 (color online): *d* and *f* LDOS curves inside the muffin-tins for the Pu atoms on the surface layer and *p* LDOS curves for O adatom. At the top site, O sits on-top of Pu1. Vertical line through E=0 is the Fermi level. LDOS correspond to calculations with SOC.

40